# Query Expansion Strategy based on Pseudo Relevance Feedback and Term Weight Scheme for Monolingual Retrieval

Rekha Vaidyanathan
Dept. of Computer Applications
MANIT, Bhopal

Sujoy Das
Dept. of Computer Applications
MANIT, Bhopal

Namita Srivastava
Dept. of Computer Applications
MANIT, Bhopal

## ABSTRACT
Query Expansion using Pseudo Relevance Feedback is a useful and a popular technique for reformulating the query. In our proposed query expansion method, we assume that relevant information can be found within a document near the central idea. The document is normally divided into sections, paragraphs and lines. The proposed method tries to extract keywords that are closer to the central theme of the document. The expansion terms are obtained by equi-frequency partition of the documents obtained from pseudo relevance feedback and by using tf-idf scores. The idf factor is calculated for number of partitions in documents. The group of words for query expansion is selected using the following approaches: the highest score, average score and a group of words that has maximum number of keywords. As each query behaved differently for different methods, the effect of these methods in selecting the words for query expansion is investigated. From this initial study, we extend the experiment to develop a rule-based statistical model that automatically selects the best group of words incorporating the tf-idf scoring and the 3 approaches explained here, in the future. The experiments were performed on FIRE 2011 Adhoc Hindi and English test collections on 50 queries each, using Terrier as retrieval engine..

## General Terms
Information Retrieval

[1]This work was presented as a poster in the FIRE 2012 Conference with only 2 methods for comparison - using only Hindi Collection for the experiment. At the time the reported performance for Hindi collection was 2.15%. In this paper the same methodologies and one more additional method is applied on Hindi and additionally English collection. The detailed analysis and observations are also explained in this paper. Here, the nature and length of the query is studied to test its impact on the proposed strategy. (Reference 1 RekhaVaidyanathan et.al 2012.)

## 1. INTRODUCTION
The purpose of query expansion is to reformulate a query by introducing terms that are closely related to original query [1]. Moreover, the queries fired by a user, typically in web search, are often short and do not provide much hint [2]. A document retrieved can be relevant even if it does not contain the words that are exactly matching with the query but is conceptually same. Thus, the most challenging task is to make the machine automatically understand the query and produce related words that are close to the original query. In literature, various techniques have been proposed by researchers. Psuedo Relevance Feedback (PRF) is one such technique where the user submits a query and the search engine retrieves the relevant documents using some retrieval method. This is sometimes called as local analysis and is treated as a special case of relevance feedback [3][4]. The top N retrieved documents are then used to reformulate the query using additional terms, apart from original query terms, for better results. Given a query and a set of relevant documents, it has been observed that there are words in a section or paragraph that are contextually related to the keywords. The question remains as how to select these. The proposed method tries to first identify that section of a document which is contextually closer to the topic given by the user. From here, meaningful words are identified for query expansion. The method employs equi-frequency partition of the top 10 documents retrieved initially using Pseudo Relevance feedback, to find the section that is contextually closer to the query or topic. Tf-idf scoring of words for that section is computed for each document. The idf part of the score is computed for each partition of the document than whole collection. The scores thus obtained are sorted in descending order to determine the selection of candidate words. An attempt is made to study how the words selected using their scores affect the query.

Three ways are explored in this paper to search these words in a retrieved document. From the list with words and their scores in descending order, the following methods are adopted to select group of words for expansion.

•  Method 1: Group of words with highest tf-idf score: Assuming that the words with highest score are related to the query.

•  Method 2: Group of words with average score in the sorted list of scores and words. The assumption here is that there is a general tendency for scores to move towards average.

•  Method 3: Group of words that contain keyword(s) having highest score is present in this group.

The terms are thus selected in 3 ways, using Method 1,2 and 3 and query is expanded using these terms to investigate how they perform.

## 2. RELATED WORKS
Automatic query expansion is difficult to perform as reformulation may cause query to drift. Normally, tf-idf is measured to extract keywords that appear frequently in a document [5]. It is observed that two words that are near to each other in a set of documents may possess some semantic relationship [6]. The position and frequency of words in a document are used to calculate their inter-quartile ranges to get the dispersion of the words within a document using Fourier series expansion [6][7]. In an extension of tf-idf weighting using Fuzzy method it was observed that the proposed weighting method produced ranking that matched human judgment [8]. Jones Sparck proposed a method where words that co-occur were clustered and used for automatically expanding the query[9]. However, Helen & Willet stated that words that co-occur very frequently can fail to discriminate between relevant and non-relevant documents [10]. Luhn stated that the frequency of occurrence of word help in measuring its significance. His method counted the words and they were ranked based on two cut off points that followed Zipf's law: Words that are too frequent beyond the upper cut-off, were not considered and words that are infrequent, below lower cut-off were rejected. Also, the relative position of a word within a sentence can also be a factor in measuring the significance of a sentence [11] .In our previous paper , frequency and position of query terms was used effectively to find the densest region of a document that has the most relevant terms to that of the query terms[12]. The documents were partitioned in two ways: equi-width and equi-frequency. An overlap of the range of words i) which has the maximum keywords and ii) the least range that has exactly k number of keywords is regarded as the densest region or the most relevant region of the document, given a query.

Robertson mentioned that the inverse document frequency or idf, proposed by Sparck Jones in 1971, counts the number of documents being searched that contains the query term [13]. In our method, we propose a variation of idf by counting the number of partitions in the document, to be coupled with term frequency.

## 3. METHODOLOGY

In this model, expansion terms are obtained by combining pseudo relevance feedback and equi-frequency partition with tf-idf scoring. After the initial retrieval, the top 10 documents are partitioned and terms are weighted using tf-idf before reformulating the query. Each document is partitioned into sections based on equi-frequency model where the frequency of the keyword in each section forms the criterion for partition. The range (or position of the words) where the keywords appear maximum is noted. Each of the top 10 documents is then partitioned based on equi-frequency model where a partition or section must have exactly k number of keywords. The calculation is given in Equation 1. Based on these partitions, the words are given scores. While calculating tf-idf score, partitions obtained through equi-frequency partition is considered as a complete entity instead of retrieved document. Thus the idf formula is changed for partitions than documents to obtain the score of words within a document. A sorted list of words based on this score is generated. In Method-1, a group of words that has highest score is selected for expansion while Method 2 , the group of words that has the average score is selected. In Method-3, group of words that have same score as that of the query term(s) (highest score among the query terms) is selected for expansion. The comparative study shows that queries behave differently for each method and this calls for a unified model that effectively uses and selects the method automatically depending on nature and size of the query.

### 3.1 Definitions

Equi-width partitioning: In equi-width or equal-width partition, the partition/bins are of equal size where the interval range of values in each bin is constant.

For a document, divide the total words into bins or partitions where each bin contains equal number of words. The range of the bin is the position of words. Usually, the number of bins are determined by calculating median, quartile, decile or percentiles.

Equi-frequency partitioning: In equi-frequency or equal-frequency partition, the bins contain exactly k values. Here, the size of the bin varies but k remains constant. The value k is derived from any formula.

For a document, divide the total words of a document in such a way that the frequency of the keyword is exactly k in number in each bin or partition. So, each bin will contain exactly k number of keywords (can be same keyword appearing more than once).Here the number of words in each bin will vary.

### 3.2 Algorithm

Automatic Query Expansion using Method 1, 2 & 3

**Input:** Top 10 Retrieved Documents based on 50 Topics/ Collection.

**Output:** Expanded Topic files for 50 Topics/Collection.

**repeat**

1. Perform equi-width partitioning of the initially retrieved top 10 document into deciles. *(di is referred as a partition of a Document D) after removing the stopwords.*

    a. *for* each partition *di*, calculate total frequency of query terms falling within that partition.

    b. Obtain the frequency of the keyword that appears maximum number of times. (Set it to fscoremax).

2. .Partition each top 10 retrieved retrieved document using equi-frequency partitioning. The frequency is a derived constant k, calculated using formula,

$$k = \frac{\sum fi}{f_{scoremax}} \quad (1)$$

*where $\sum fi$ is the total frequency of keywords in the document d and $f_{scoremax}$ is frequency of the keyword that appeared maximum number of times from equi-width partition. We assume, k, is such that any region of a document will have exactly k number of keywords.*

3. Assuming that each Partition is equal to a document in the original formula, the tf-idf is calculated for each partition.

$$t(f, p) = \frac{f(t, p)}{\max\{f(w, p): w \in p\}} \quad (2)$$

*where t(f,p) is the frequency of a term in partition p, max {f(w,p)} is the maximum frequency in partition p*

$$\text{idf}(t, P) = \log\left[\frac{P}{\{p \in P: t \in p\}}\right] \quad (3)$$

*P = total number of partitions ; {p∈P: t∈p}= number of partitions where term t appears*

$$\text{Score}(t, p, P) = t(f, p) * \text{idf}(t, p) \quad (4)$$

4. Select the Top 5 words and store in a list.

5. Repeat Steps 1,2,3,4 for top 10 documents.

6. Sort the final list of words in descending order of Scores for each topic and write it into a keywords file.

7. Extract additional terms or expansion words from this sorted list, using the following 3 methods, for reformulating the query.

   a) **Method-1**: *Select group of words with maximum Score as expansion words. (This group contains top 5 words. This may or may not contain keywords)*

   $$G_{di} = Max\,(Score(t,p,P))$$

   b) **Method-2**: *Select Group of words having average Score.*

   $$G_{di} = Average(Score(t,p,P))$$

   c) **Method-3**: *Select Group of words that has maximum number of keywords in the sorted score list. If there is only one keyword appearing per group, the group with highest score is chosen. If no keyword appears in the list, we do not expand the query q.*

   $$G_{di} = Score(t,p,P)\ where\ t\ \epsilon\ q$$

8. Resubmit the query using the expansion words selected with these methods.

## 3.3 Experiment and Discussion

The experiment is performed on FIRE 2011 adhoc Hindi and English test collections, using Terrier retrieval engine. The query is formulated using "title" field of FIRE 2011 Hindi and English test collection. Both the collections typically contain News from newspapers and the format of topic files adheres to the TREC specifications. Each collection contains 50 topics.

For our evaluation purpose, 50 queries for each English and Hindi collection were fired. An experiment on FIRE 2011 Adhoc Hindi data is referred as HTRUN and Adhoc English data is referred as ETRUN.

A Sample of the expanded terms obtained along with scores using the three methods is shown in Table 1 for topic no. 138 for query <title> भारत में महिला आरक्षण बिल </title>

**Table 1. Group Score and Per Query Score for the three methods**

| Method | Group Score | Expansion words | Per Query Score (After Expansion) |
|---|---|---|---|
| 1.Highest score | 1.0 | गांवों, पुराना, खड़ा, समर्थन, तहत | 0.1025 |
| 2. Average Score | 0.477 | स्वरूप,अनुसूचित,मौजूदा,जेपीसी,बारे, चर्चा,मिला,प्रस्ताव, फैसला,सकती संसद,सत्र | 0.0393 |
| 3.Keyword score | 0.397 | महिलाओं, कांग्रेस, पास, बजट | 0.1759 |

Though per-query score difference may not statistically significant to make a generalization, it is observed that Method 3 has an edge over the other two methods. The average score has some of the drifting words that reduce the Per Query Score significantly. The number of words fetched also deteriorates the performance of the query.

Table 2 shows the Mean Average Precision obtained for HTRUN after conducting the experiments using the three methods.

**Table 2. Comparison of Average Precision incase of before and after Query Expansion using Method 1, 2, 3 for HTRUN**

| | Before Expansion | Highest Score | Keyword Score | Average Score |
|---|---|---|---|---|
| No.Of Queries | 50 | 50 | 50 | 50 |
| Retrieved | 49907 | 50000 | 50000 | 50000 |
| Relevant | 2885 | 2885 | 2885 | 2885 |
| Relevant Retrieved | 2059 | 1736 | 2050 | 1857 |
| **Average Precision** | **0.2453** | **0.1895** | **0.2507** | **0.1934** |

It is observed that group of words with Highest Score and Average Score resulted in lower mean average precision as compared to group that has the keyword. The query was improved by 2.15 %.

Table 3 shows the Mean Average Precision for ETRUN on 50 topics using the three methods.

**Table 3. Comparison of Average Precision incase of before and after Query Expansion using Method 1, 2, 3 for ETRUN**

| | Before Expansion | Highest Score | Keyword Score | Average Score |
|---|---|---|---|---|
| No. Of Queries | 50 | 50 | 50 | 50 |
| Retrieved | 50000 | 50000 | 50000 | 50000 |
| Relevant | 2761 | 2761 | 2761 | 2761 |
| Relevant Retrieved | 2314 | 2166 | 2201 | 2229 |
| **Average Precision** | **0.2965** | **0.2782** | **0.2650** | **0.2762** |

It is observed that none of the methods improved the MAP in case of English corpus.

The per-query analysis shows that each of these methods performed well for some or the other queries. Fig 1& 2 show the percentage of queries that improved with the three methods in 50 Hindi and 50 English queries.

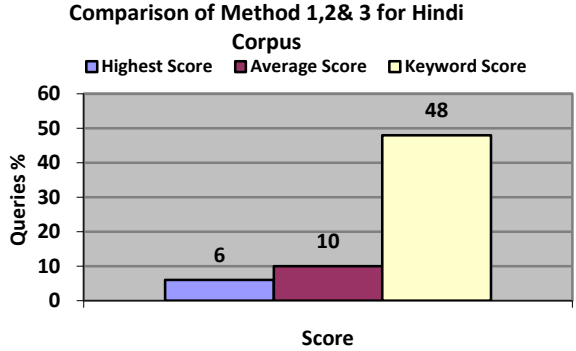

**Fig 1. Percentage of queries that improved with each of the three different methods in Hindi Corpus**

It is clear that Keyword Score group improved more queries compared to the other two methods. For Hindi collection, 3 in 50 queries improved with Highest Score. 5 in 50 queries showed improvement with Average Score and 24 queries improved with Keyword Score group. Examples for each of the methods that improved the precision are given below.

**Method 1**: Highest Score performed well for topic 133 <title>अबू गरीब जेल में अत्याचार </title>

**Table 4. Group Score and Per Query Score for the three methods –Topic 133 Hindi**

| Method | Group Score | Expansion words | Per Query Score (After Expansion) |
|---|---|---|---|
| 1.Highest score | 0.845 | **इलाज, सौ, नसीब, प्रतिदिन** | 0.2848 |
| 2.Average Score | 0.389 | वाला, चर्चा | 0.1855 |
| 3.Keyword score | 0.602 | प्रधानमंत्री, अमेरिकी, वीटो, स्थाई, मुठभेड़ | 0.1759 |

**Method 2** : Average Score performed well for topic 146 <title> राम जन्मभूमि मामले का फैसला <title>

**Table 5. Group Score and Per Query Score for the three methods –Topic 144 Hindi**

| Method | Group Score | Expansion words | Per Query Score (After Expansion) |
|---|---|---|---|
| 1.Highest score | 0.903 | मुद्दा, मुद्दे, यथास्थिति, जज, अहमद | 0.1 |
| 2.Average Score | 0.465 | **दाखिल,अपील,निर्माण** | 0.2079 |
| 3.Keyword score | 0.698 | पद्धति, संबंधित, पूरी, अपना, मामले, करा, याचिका | 0.1665 |

Method 3: Keyword Score performed well for topic 169 <title> नक्सली हमला</title>

**Table 6. Group Score and Per Query Score for the three methods –Topic 169  Hindi**

| Method | Group Score | Expansion words | Per Query Score (After Expansion) |
|---|---|---|---|
| 1.Highest score | 0.698 | २००६, छत्तीसगढ़, भाग, बलों, खिलाफ, बंगाल, गुनी, पश्चिम, रमन, सिंह | 0.1382 |
| 2.Average Score | 0.225 | आदिवासियों, मुठभेड़ | 0.1274 |
| 3.Keyword score | 0.357 | नक्सलियों | 0.3337 |

Given in Fig 2 is the percentage of queries that were improved using the three methods for English Corpus.

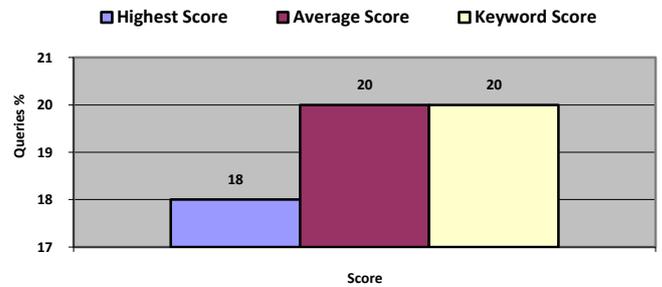

**Fig 2. Percentage of queries that improved with each of the three methods in English Corpus**

From Fig 2, we can see that Group with Average and Keyword score improved precision more than the Highest Score, though the difference is not very highly significant. 9 in 50 Queries improved with Highest Score,10 each for both Average and Keyword Score for English.

In English, 1/11 th of Highest Score group had a keyword group in them. 1/6th of Average score group had a keyword group in them. Thus it can be said that the presence of keyword in a group does make some difference compared to the group in which query words or keywords are not there in spite of their higher score. It is a matter of interest as to why English topics fared better compared to the Hindi topics in each of the three methods. Examples of the topics that improved with each of the methods and the expansion words obtained are given below:

**Method 1:** Topic that improved with Highest Score : topic no: 139 <title>Vanquishing the Somali pirates</title>

**Table 7. Group Score and Per Query Score for the three methods –Topic 139 English**

| Method | Group Score | Expansion words | Per Query Score (After Expansion) |
|---|---|---|---|
| 1.Highest score | 0.698 | ministry, board, assets, work, egyptian, naval, grown | 0.5118 |
| 2. Average | 0.477 | year, raids, coast, khaliq east, west, international, gulf, | 0.3944 |

| | Score | | maritime, crew, traders, release | |
|---|---|---|---|---|
| 3.Keyword score | 0.602 | | million,supertanker,mogadishu, samho ,nov., ship, seize, meeting , problem, tried, accused, piracy, evidence, firm, mother, pirates operations, others | 0.4684 |

**Method 2**: Topic that improved with Average Score : topic no: 127

<title> Rare cosmic events</title>

**Table 8. Group Score and Per Query Score for the three methods –Topic 127 English**

| Method | Group Score | Expansion words | Per Query Score (After Expansion) |
|---|---|---|---|
| 1.Highest score | 0.698 | organisations,river, scientific,event, hooghly,hundreds | 0.0401 |
| 2. Average Score | 0.318 | view,still,absolutely | 0.1439 |
| 3.Keyword score | 0.698 | organisations,river, scientific,event, hooghly,hundreds | 0.0401 |

Method 3: Topic that improved with Keyword

Score : topic no: 174

<title> International economic slump </title>

**Table 8. Group Score and Per Query Score for the three methods –Topic 174 English**

| Method | Group Score | Expansion words | Per Query Score (After Expansion) |
|---|---|---|---|
| 1.Highest score | 0.698 | sekhar,calcutta,b-school, performances, external, krishnan | 0.0005 |
| 2. Average Score | 0.477 | eported,reserve,economy,opec,Thursday, u.s. | 0.0092 |
| 3.Keyword score | 0.602 | alarming,fund, international, monetary,asian, development | 0.0233 |

### 3.3.1 Observations:
*On Length of the Query*

> In Hindi Collection, short group of words gave better results.

> In very few queries large number of expanded words as large as 14 showed significant improvements in Average Precision. This calls for a unified model that can identify the best group from the sorted list of words, irrespective of whether they fall in the highest score, Average score or Keyword score group.

*Nature of Query*

> Groups with nouns,plurals and classifications performed well.

> It is noted that the presence of the keyword in the group did make a difference but it is also affected by the number of words and the meaning.

Table 3 shows how for topic 163 of hindi test collection, titled

"भारत में बाघ संरक्षण" ("Tiger conservation in India") ; a plural and the classification of tiger did bring about a better result than words like tiger project, law etc which are generic in nature but may be contextually correct.

**Table 9. Group Score and Per Query Score for the three methods –Topic 163Hindi**

| Words | Precision |
|---|---|
| टाइगर, प्रोजेक्ट, कानून, भारतीय, बाघ, संरक्षण, परियोजना, संशोधन, प्राधिकरण, शुरू, धरोहर | 0.2719 |
| बाघों, वन्यजीवों | 0.3513 |

In such cases, the typical problem of finding an ideal query length remains to be explored as longer queries did hurt precision.

> In English Collections, the average group performed better than others. This gives us a clue to the effect that the distribution pattern of the score may have a role selecting the group of words for expansion.

> The three methods considered are fairly simple in nature and can thus be expanded to form a statistical model that automatically selects the right group of words based on their score. Also, the length of the query needs to be considered to avoid query drift.

## 4. CONCLUSION

In this paper, expansion terms are obtained by combining pseudo relevance feedback and equi-frequency partition with tf-idf scoring technique. While calculating **tf-idf** score, partition is considered as a complete entity instead of the retrieved document. The experimental results show that the group of words that have the same tf-idf score as that of query terms did bring about some effect in selecting the group of words for expansion.Also, for some topics the average scores performed better. Further, merely taking the highest scored words based on tf-idf, after partitioning the initial top 10 retrieved documents, based on equi-frequency partition and tf-idf may not necessarily improve retrieval effectiveness. It is most important to note that the following factors do affect the precision of any augmented query.

- Length of the augmented query
- Nature of words (synonym,plural,classifiers etc) and
- The distribution pattern of the words and their scores.

We are carrying out further study and analysis to improve the retrieval performance by modeling the same by considering the above factors to formulate a statistical model/framework for automatically selecting the words for expansion.

## 5. ACKNOWLEDGMENTS

Our thanks are due to Terrier™ team for providing free software for researchers in the field of IR [Iadh Ounis et.al, 2006]. Our sincere thanks to FIRE group allowing us to use the data for our experiment.